\documentclass[floatfix, preprint, showpacs, showkeys, preprintnumbers, nofootinbib, superscriptaddress]{revtex4-1}
\usepackage{comment}
\usepackage{amssymb}
\usepackage{color}
\usepackage{rotating}
\usepackage{amsmath}
\usepackage{ulem}
\usepackage{overpic}
\usepackage{graphicx}
\usepackage{dcolumn}
\usepackage{graphicx}
\usepackage{bm}
\usepackage{chngpage}
\usepackage{multirow}
\usepackage{slashed}
\usepackage{indentfirst}
\usepackage{booktabs}
\usepackage{amssymb,amsfonts}
\usepackage{amsmath}
\usepackage{color}
\usepackage[colorlinks,citecolor=blue,linkcolor=blue,hypertex,breaklinks=true]{hyperref}

\begin{document}
\title{Odd-even staggering and shell effects of charge radii for nuclei with even $Z$ from $36$ to $38$ and from $52$ to $62$}

\author{Rong An}
%\email[]{anrong@brc.ac.cn}
%\affiliation{Key Laboratory of Beam Technology of Ministry of Education, Beijing Radiation Center, Beijing 100875, China}
\affiliation{Key Laboratory of Beam Technology of Ministry of Education, Institute of Radiation Technology, Beijing Academy of Science and Technology, Beijing 100875, China}
\affiliation{Key Laboratory of Beam Technology of Ministry of Education,
College of Nuclear Science and Technology, Beijing Normal University, Beijing 100875, China}

\author{Xiang Jiang}
%\email[]{jiangxiang@szu.edu.cn}
\affiliation{College of Physics and Optoelectronic Engineering, Shenzhen University, Shenzhen 518060, China}

\author{Li-Gang Cao}
%\email[]{caolg@bnu.edu.cn}
\affiliation{Key Laboratory of Beam Technology of Ministry of Education, College of Nuclear Science and Technology, Beijing Normal University, Beijing 100875, China}
\affiliation{Key Laboratory of Beam Technology of Ministry of Education, Institute of Radiation Technology, Beijing Academy of Science and Technology, Beijing 100875, China}

\author{Feng-Shou Zhang}
\email[Corresponding author: ]{fszhang@bnu.edu.cn}
\affiliation{Key Laboratory of Beam Technology of Ministry of Education, Institute of Radiation Technology, Beijing Academy of Science and Technology, Beijing 100875, China}
\affiliation{Key Laboratory of Beam Technology of Ministry of Education, College of Nuclear Science and Technology, Beijing Normal University, Beijing 100875, China}
\affiliation{Center of Theoretical Nuclear Physics, National Laboratory of Heavy Ion Accelerator of Lanzhou, Lanzhou 730000, China}

%\date{\today}

\begin{abstract}
 A unified theoretical model reproducing charge radii of known atomic nuclei plays an essential role in making extrapolations for unknown nuclei. Recently developed new ansatz which phenomenologically takes into account the neutron-proton short-range correlations ($np$-SRCs) can describe the discontinuity properties and odd-even staggering (OES) effect of charge radii along isotopic chains remarkably well. In this work, we further review the modified root-mean-square (rms) charge radii formula in the framework of relativistic mean field (RMF) theory. The charge radii are calculated along various isotopic chains that include the nuclei featuring the $N=50$ and $82$ magic shells. Our results suggest that RMF with and without considering a correction term give an almost similar trend of nuclear size for some isotopic chains with open proton shell, especially the abrupt increases across the strong neutron closed shells and the OES behaviors. This reflects that the $np$-SRCs have almost no influence for some nuclei due to the strong coupling between different levels around Fermi surface. The weakening OES behavior of nuclear charge radii is observed generally at completely filled neutron shells and this may be proposed as a signature of magic indicator.
\end{abstract}

%\pacs{25.70.Jj,24.10.-i}

\maketitle
\section{Introduction}
%%%% experiments
Nuclear charge radii, which can characterize the charge density distributions and the Coulomb potentials in nuclei, provide access to nuclear structure information~\cite{Angeli_2009}. Plenty of methods are employed to perform measurements of nuclear charge radii, such as muonic atom x rays ($\mu^{-}$)~\cite{BAZZI2011199}, high-resolution laser spectroscopy techniques~\cite{PhysRevLett.93.113002,Cheal_2010,Blaum_2013,vernon2020laser,PhysRevLett.125.023002,PhysRevLett.108.142501}, high energy elastic electron scattering ($e^{-}$)~\cite{PhysRevC.21.1426,Avgoulea_2011,PhysRevLett.102.102501} and isotope shifts (ISs)~\cite{Manovitz2019PRL,OZAWA200132}, etc. So far, more available charge radii data are provided in the nuclear chart~\cite{ANGELI201369}. As one of the important input quantities in astrophysics, nuclear charge radius plays an important role in theoretical study~\cite{ARNOULD2020103766}. Moreover, reliable predictions can also serve as useful guides for experimental detection of charge radii of nuclei far away from the $\beta$-stability line.

%%%%introduce the specific features of charge radii

In general, nuclei charge radii often display the emergence of simple patterns and regular behaviours or global properties that the variations of charge radii along isotopic chains represent discontinuous features~\cite{ANGELI201369,GarciaRuiz:2019cog,PhysRevC.100.044310,PhysRevC.90.054318}. These remarkably abrupt changes in charge radii are observed naturally across the neutron-closure shells, namely the kinks at $N=20$, $28$, $50$, $82$, $126$~\cite{ANGELI201369,Ruiz2016,deGroote,Koszorus:2020mgn,PhysRevC.100.034304,Hammen2018,Gorges2019,PhysRevLett.110.032503,goodacre2020laser,PhysRevC.65.054320}. The strong shell structure results in the parabolic-like shapes of charge radii with respect to the variation of neutron number. In addition, the odd-even staggering (OES) effect that the nuclear charge radii of odd-$N$ isotopes are smaller than the averages of their even-$N$ neighbors, is generally observed throughout the nuclear chart~\cite{ANGELI201369}.

%%%% empirical formula and theoretical models (negative)

With the accumulation of experimental data, many empirical relations and microscopic models have been proposed to investigate the variations of nuclear charge radii. The general nuclear size is ruled by $A^{1/3}$ law through introducing the shell and isospin effects~\cite{PhysRevC.88.011301,Sheng:2015poa}. Similarly, $Z^{1/3}$ dependence for nuclear charge radius is also directly proposed in Ref.~\cite{Zhang:2001nt}. The sophisticated Garvey-Kelson (GK) relation had been transformed to describing the nuclear charge radius~\cite{PhysRevC.94.064315,Piekarewicz2010}, but its extrapolating ability is limited~\cite{PhysRevC.84.014333,PhysRevC.89.061304}. For heavy or superheavy neutron-rich elements, it is worth noting that their bulk properties are barely obtained due to short half-lives~\cite{zhang2018}. Recent works attempted to deduce the charge radii based on the $\alpha$-decay properties~\cite{PhysRevC.89.024318,PhysRevC.87.024310}, even cluster and proton emission data~\cite{PhysRevC.87.054323}. The microscopic nuclear structure models based on the mean-field approach such as Hartree-Fock-Bogoliubov (HFB) model~\cite{PhysRevLett.102.242501,PhysRevC.82.035804} and relativistic mean field (RMF) theory~\cite{Geng:2003pk,PhysRevC.82.054319} can reveal the inner nuclear interactions self-consistently. As encountered in $ab~initio$ calculations with chiral effective field theory (EFT) interactions~\cite{Ruiz2016}, these models cannot reproduce the fine structure of nuclear charge radii well. In recent years, Bayesian neural networks as an alternative approach were devoted to describe the charge radii in the nuclear chart~\cite{Utama_2016,RenZZ2020,Wu:2020bao}. This approach can describe well the experimental data within uncertainties, but the underlying physical mechanism is unclear.

%%%%(positive) methods
Various mechanisms have been proposed to elucidate the evolution of nuclear charge radii, such as the core polarization by valence neutrons~\cite{CAURIER198015,TALMI1984189}, precise knowledge of radial moments~\cite{Reinhard2020PRC}, adjacent nucleons relations~\cite{Bao2020PRC}, quadrupole deformation~\cite{An:2021rlw}, etc. The sophisticated Fayans EDF model, in which a novel density-gradient term was introduced into the pairing interaction, can reproduce the staggering effects of charge radii for Ca and Sn isotopes~\cite{Reinhard2017,Gorges2019}. This model demonstrates that surface pairing components play an essential role. By contrast, the phenomenologically modified charge radii formula which associates to the neutron-proton ($np$) pair correlations was proposed in the relativistic mean field (RMF) theory within NL3 parametrization set~\cite{An:2020qgp}. This new ansatz can remarkably describe the discontinuity properties and OES effects of charge radii along Cu and In isotopic chains~\cite{an2021discontinuity}. As argued in Ref.~\cite{Miller:2018mfb}, the short-range correlations (SRCs) originating from $np$ pairing contribute to the root-mean-square (rms) charge radius of finite nuclei. The $np$ correlations play an essential role in characterizing the OES of nuclear charge radii~\cite{Zawischa:1985qds} and the magnitude of the neutron skin in asymmetric nuclei~\cite{Ryckebusch2021}.

%%%% goal of this paper

 As mentioned in Ref.~\cite{Miller:2018mfb}, the effect of $np$-SRCs has an influence on the computed rms charge radius. In some cases, however, this effect has no contributions to the charge radius. In Ref.~\cite{goodacre2020laser}, it pointed out that pairing did not play a crucial role in the origin of the kink at magic number and OES behaviors. This means the experimental data can be reproduced predominately at the mean-field level. To understand these uncertainties, we further check the recently developed approach by studying the charge radii of nuclei featuring the $N=50$ and $82$ magic shells. We focus on the variations of nuclear charge radii and the OES behaviors along isotopic chains and aim to make a further complement for our new ansatz.

This paper is organized as follows. In Sec.~\ref{second}, we briefly report the theoretical model. In Sec.~\ref{third}, we present the results and discussions. A short summary and outlook are provided in Sec.~\ref{fourth}.

\section{Theoretical model}\label{second}
The relativistic mean field (RMF) theory had made remarkable successes in describing various nuclear physics phenomena~\cite{Vretenar:2005zz,Meng:2005jv, Liang:2014dma, jie2016relativistic,PhysRevC.67.034312,PhysRevC.69.054303,zhang2007,PhysRevC.90.044305,zhang2012,PhysRevC.92.024324,Cao:2003yn}. For nonlinear self-consistent Lagrangian density, nucleons are described as Dirac particles which interact via the exchange of $\sigma$, $\omega$ and $\rho$ mesons. The electromagnetic field is served as photon. The effective Lagrangian density is written as
\begin{eqnarray}
\mathcal{L}&=&\bar{\psi}[i\gamma^\mu\partial_\mu-M-g_\sigma\sigma
-\gamma^\mu(g_\omega\omega_\mu+g_\rho\vec
{\tau}\cdotp\vec{\rho}_{\mu}+eA_\mu)]\psi\nonumber\\
&&+\frac{1}{2}\partial^\mu\sigma\partial_\mu\sigma-\frac{1}{2}m_\sigma^2\sigma^2
-\frac{1}{3}g_{2}\sigma^{3}-\frac{1}{4}g_{3}\sigma^{4}\nonumber\\
&&-\frac{1}{4}\Omega^{\mu\nu}\Omega_{\mu\nu}+\frac{1}{2}m_{\omega}^2\omega_\mu\omega^\mu
+\frac{1}{4}c_{3}(\omega^{\mu}\omega_{\mu})^{2}-\frac{1}{4}\vec{R}_{\mu\nu}\cdotp\vec{R}^{\mu\nu}\nonumber\\
&&+\frac{1}{2}m_\rho^2\vec{\rho}^\mu\cdotp\vec{\rho}_\mu
+\frac{1}{4}d_{3}(\vec{\rho}^{\mu}\vec{\rho}_{\mu})^{2}-\frac{1}{4}F^{\mu\nu}F_{\mu\nu},
\end{eqnarray}
where $M$ is the mass of nucleon, $m_{\sigma}$, $m_{\omega}$ and $m_{\rho}$ are the masses of the $\sigma$, $\omega$ and $\rho$ mesons, respectively. Here $g_{\sigma}$, $g_{\omega}$, $g_{\rho}$ and $e^{2}/4\pi$ are the coupling constants for $\sigma$, $\omega$, $\rho$ mesons and photon, respectively. In the present work, the Dirac equation for the nucleons and the Klein-Gordon type equations with sources for the mesons and the photon are solved by the expansion method with the axially symmetric harmonic oscillator basis~\cite{Geng:2003pk}. Twelve shells are used for expanding the fermion fields and 20 shells for the meson fields. The NL3 parameter set is employed~\cite{PhysRevC.55.540}. In order to obtain the ground state properties, the Hamiltonian of the system becomes $H'=H-\lambda\langle{Q}\rangle$, the second term on the right hand represents the modified linear constraint part~\cite{PhysRevC.89.014323}, where $\lambda$ is the spring constant that its value is changed during the self-consistent iteration and $Q$ is intrinsic multipole moment. The ground state properties of finite nuclei are obtained by constraining quadrupole deformation $\beta_{20}$. The values of $\beta_{20}$ change from $-0.60$ to $0.60$ with the interval of $0.01$.
In general, the mean-square charge radius of a nucleus has the form (in units of fm$^{2}$)~\cite{Ring:1997tc,Geng:2003pk}
\begin{eqnarray}\label{coop1}
R_{\mathrm{ch}}^{2}=\frac{\int{r}^{2}\rho_{\mathrm{p}}(\mathbf{r})d^{3}r}{\int\rho_{\mathrm{p}}(\mathbf{r})d^{3}r}+0.64~\mathrm{fm}^2,
\end{eqnarray}
where in the first term, $\rho_{\mathrm{p}}(\mathbf{r})$ corresponds to the charge density distribution of point-like proton and then the second term accounts for the finite size effects of proton~\cite{Ring:1997tc}. However, the shell closure and OES effect of charge radii cannot be reproduced well for Ca isotopes~\cite{An:2020qgp}. Since the modified expression had been proposed in the following way:
\begin{eqnarray}\label{coop2}
R_{\mathrm{ch}}^{2}=\frac{\int{r}^{2}\rho_{\mathrm{p}}(\mathbf{r})d^{3}r}{\int\rho_{\mathrm{p}}(\mathbf{r})d^{3}r}+0.64~\mathrm{fm}^2+\frac{a_{0}}{\sqrt{A}}\Delta{{D}}~\mathrm{fm}^2.
\end{eqnarray}
The last term on the right hand is the modified term which associates to the Cooper pair condensation~\cite{PhysRevC.76.011302}. The quantity $A$ is the mass number and $a_{0}=0.834$ is a normalization constant for nuclei with $Z<50$. For nuclei with $Z\geq50$, $a_{0}=0.22$ is used~\cite{An:2020qgp}. The quantity $\Delta{D}=|{D}_{\mathrm{n}}-{D}_{\mathrm{p}}|$ represents the difference of Cooper pair condensation between neutron and proton.
In Eq.~(\ref{coop2}), the modified term is associated to the Cooper pair condensation function with the following form:
\begin{eqnarray}\label{coop3}
{D}_{\mathrm{n,p}}=\sum_{k>0}^{\mathrm{n,p}}u_{k}v_{k}.
\end{eqnarray}
This quantity can represent a measurement of the number of Cooper pairs in the BCS wave function~\cite{PhysRevC.76.011302}, where $v_{k}$ and $u_{k}$ are the BCS amplitudes of occupation and non-occupation probability of the $k$th single-particle orbital, respectively. The summation is over all the occupied single-particle levels. It is calculated self-consistently by solving the state-dependent BCS equations with a $\delta$ force interaction~\cite{GENG200480,20181107}. More details and discussions are shown in Ref.~\cite{An:2020qgp}.

\section{Results and discussion}\label{third}
\subsection{Cooper pair components}
As mentioned above, the information about Cooper pair components is obtained by tackling the pairing correlations in atomic nuclei. Conventionally, the pairing correlations can be treated either by the BCS method or by the Bogoliubov transformation. The Bogoliubov transformation is widely used in a theoretical framework, such as in the great successful relativistic Hartree-Bogoliubov (RHB) models~\cite{PhysRevLett.77.3963,MENG19983,PhysRevC.82.011301,PhysRevC.85.024312}. Especially, recent study shows that the deformed RHB model in continuum makes a predictive power for nuclear mass in regions of astrophysical interest~\cite{PhysRevC.104.L021301}. In like wise, the BCS method can also capture the ground state properties of finite nuclei throughout the periodic table~\cite{Geng:2003pk}. In this work, we focus on the behavior of charge radii along isotopic chains with respect to the variation of neutron number. The pairing window of 12 MeV both above and below the Fermi surface is employed for particle-particle channel. The pairing strength is generally determined by fitting to the odd-even mass staggering~\cite{Bender:2000xk}. In order to reflect the universality of our results, the pairing strength is $V_{0}=322.8$ MeV fm$^{3}$ for all calculations~\cite{An:2020qgp}. The accuracy of the convergence is determined by the self-consistent iteration in binding energy, which is lower than $10^{-6}$ MeV.

\begin{figure}[htbp]
\includegraphics[scale=0.5]{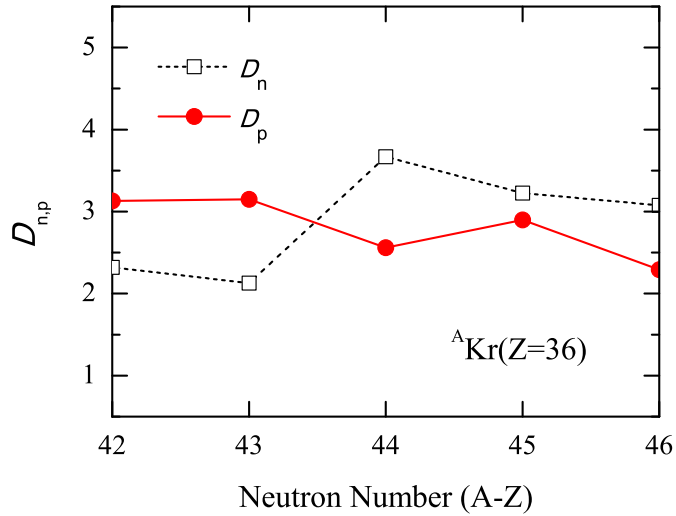}
   \caption{Cooper pair components of proton (solid circle) and neutron (open square) are shown along Kr isotopic chain.} \label{fig0}
\end{figure}
 The difference of ${D}_{\mathrm{n}}$ and ${D}_{\mathrm{p}}$ is employed to equivalently measure the $np$-SRCs components in the rms charge radius. As shown in our previous work~\cite{An:2020qgp}, the fine structure of finite nuclei charge radius is determined by the modified term. In Ref.~\cite{Miller:2018mfb} the authors demonstrated that the $np$-SRCs had no contribution to charge radii for some open shell nuclei. This can be easily understood in our new ansatz that ${D}_{\mathrm{n}}$ and ${D}_{\mathrm{p}}$ might be comparable for open shell nuclei due to the strong coupling between different levels around the Fermi surface. From this point of view, the quantities ${D}_{\mathrm{n,p}}$ as a function of neutron number are plotted along Kr~$(Z=36)$ isotopic chain in Fig.~\ref{fig0}. One can find the values of quantity ${D}_{\mathrm{n}}$ and ${D}_{\mathrm{p}}$ are indeed close for neutron and proton. Consequently, the modified part has slight influence on the rms charge radii. In order to clarify this phenomenon, more isotopic chains are investigated.

\subsection{Variations of charge radii along isotopic chains}
\begin{figure}[htbp]
\includegraphics[scale=0.6]{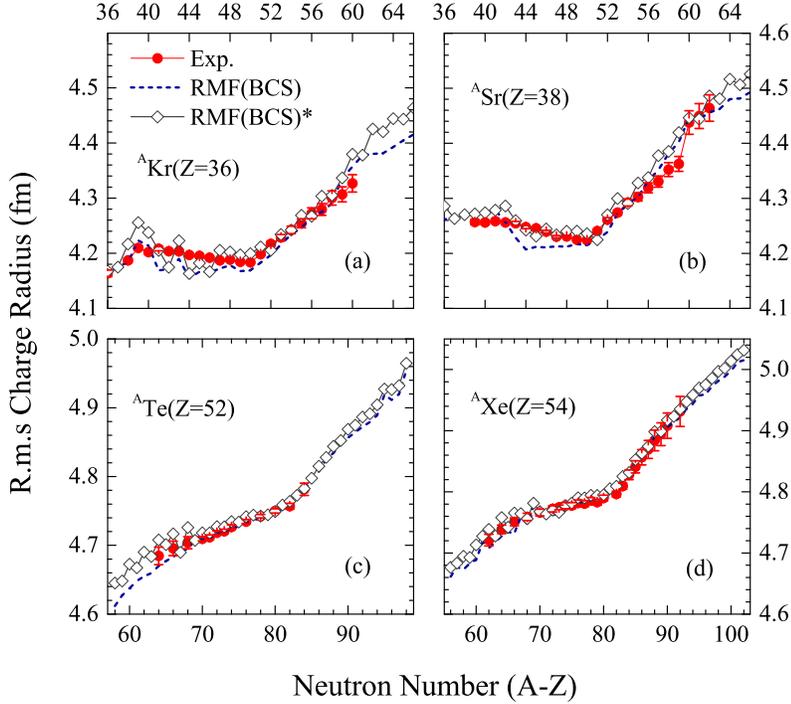}
   \caption{Charge radii of krypton (a), strontium (b), tellurium (c) and xenon (d) isotopes are obtained by the RMF(BCS) (short-dashed line) and the modified RMF(BCS)* (open diamond) approaches, respectively. The experimental data are taken from Ref.~\cite{ANGELI201369} (filled circle).} \label{fig1}
\end{figure}
For the convenience of discussion, the results obtained by Eq.~(\ref{coop1}) are labeled as RMF(BCS), and RMF(BCS)* represents the results performed by the modified charge radii formula Eq.~(\ref{coop2}). In Fig.~\ref{fig1}, charge radii of krypton (a), strontium (b), tellurium (c) and xenon (d) isotopes are obtained by the RMF(BCS) and RMF(BCS)* methods. For krypton (a) and strontium (b) isotopes, the charge radii of these two isotopic chains vary smoothly with increasing neutron number until $N=50$. The rapid increasing trends are shown across $N=50$, but the slope of the change of strontium isotopic chain is larger than krypton's. The similar rapid increase of charge radii across $N=50$ was studied earlier in RMF within NL-SH parametrization set~\cite{LALAZISSIS1995201}. These nuclei with closed shells are more difficult to excite than their neighbors, which is evidenced by their relatively stable properties~\cite{Steppenbeck2013,GarciaRuiz:2019cog,CORTES2020135071}. This suggests that these regular and rapidly increasing patterns of charge radii across $N=50$ are common features observed in self-bound many-nucleon systems~\cite{ANGELI201369}.

For $^{75}$Kr, there exists a large deviation between experimental value and the calculated result. The distinctive aspect that the shape deformation has an influence on the charge radius should be considered carefully~\cite{PhysRevC.88.011301,An:2021rlw}. The experimental data indicates the possible quadrupole deformation parameter $\beta_{20}\approx0.27$ for nucleus $^{75}$Kr~\cite{nndc}. However, in our calculation, $\beta_{20}$ is around $0.48$ for this nucleus. The same scenario can also be encountered for $^{96}$Kr, but with the calculated quadrupole deformation parameter $\beta_{20}\approx0.32$. According to the above discussions, the quadrupole deformation parameter in experiment may be much smaller than the calculated value. Along the Sr isotopic chain, the charge radii beyond the neutron number $N=60$ appear to increase with surprisingly leaping slope. As argued in Ref.~\cite{Togashi2016PRL}, some particular isotopes, as in the region around $Z=40$, $N=60$ and $Z=62$, $N=90$, are considered to present a rapid onset of deformation. Actually, for $^{98-100}$Sr, the calculated quadrupole deformation parameters $\beta_{20}$ are almost $0.45$. For $^{81-88}$Sr isotopes, the rms charge radii formula with correction term can reproduce experimental data well. But the rms charge radii of some proton-rich nuclei are slightly overestimated by the modified formula.

Remarkably, the emergency of rapidly increasing patterns of charge radii is also observed across $N=82$ in various different isotopic chains~\cite{ANGELI201369,GarciaRuiz:2019cog}. As shown in Fig.~\ref{fig1} (c) and (d), this feature is also found for tellurium and xenon isotopic chains. However, these two methods almost give similar results with the increasing neutron number.
\begin{figure}[htbp]
\includegraphics[scale=0.6]{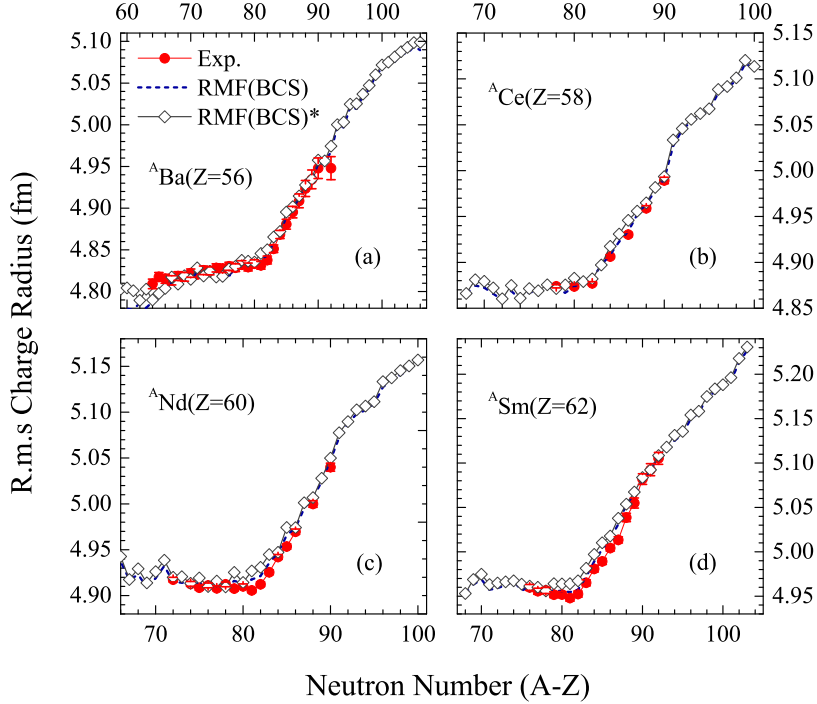}
   \caption{Same as Fig.~\ref{fig1}, but for barium (a), cerium (b), neodymium (c) and samarium (d) isotopes.} \label{fig2}
\end{figure}
In order to further elaborate the discontinuity aspects of nuclear charge radii across the $N=82$ neutron shell, in Fig.~\ref{fig2}, the calculated results for barium (a), cerium (b), neodymium (c) and samarium (d) isotopic chains are also shown. We can find that both of these two methods almost give similar trends with the increase of neutron number, especially the effect of shell closures on charge radii across $N=82$. Similar to Fig.~\ref{fig2}~(c), the RHB theory describes very well the kink at $N = 82$ in the charge radii of neodymium isotopes~\cite{PhysRevC.102.024314}.
This discontinuity aspect was presented evidently owing to the relatively stable properties of closed-shell nuclei~\cite{Hammen2018,Gorges2019}. This had been attributed to the rather small isospin dependence of spin-orbit term in RMF model~\cite{Sharma1995PRL}.

The parabolic-like shapes of nuclear charge radii are observed generally between two strong neutron closed-shells~\cite{ANGELI201369,GarciaRuiz:2019cog}. This distinct feature
was evidently presented between $N=20$ and $N=28$ shells along the calcium isotopic chain~\cite{ANGELI201369,Ruiz2016}. In addition, this peculiar phenomenon can also be found dramatically in the latest study of cadmium~\cite{Hammen2018} and tin~\cite{Gorges2019} isotopic chains. In contrast to calcium isotopes, the amplitudes of the parabolic-like shape of nuclear charge radii had been reduced remarkably.
As shown in Ref.~\cite{An:2020qgp}, the convex behavior of nuclear charge radii between two magic neutron shells can be well reproduced by the modified formula. Therefore, we may infer that the parabolic-like shape of nuclear charge radii can also be observed between $N=82$ and $N=126$ shell closures, but with smaller amplitudes. Thus more reliable experimental data are urgently needed.

\subsection{OES behaviors in nuclear charge radii}
\begin{figure}[htbp]
\includegraphics[scale=0.6]{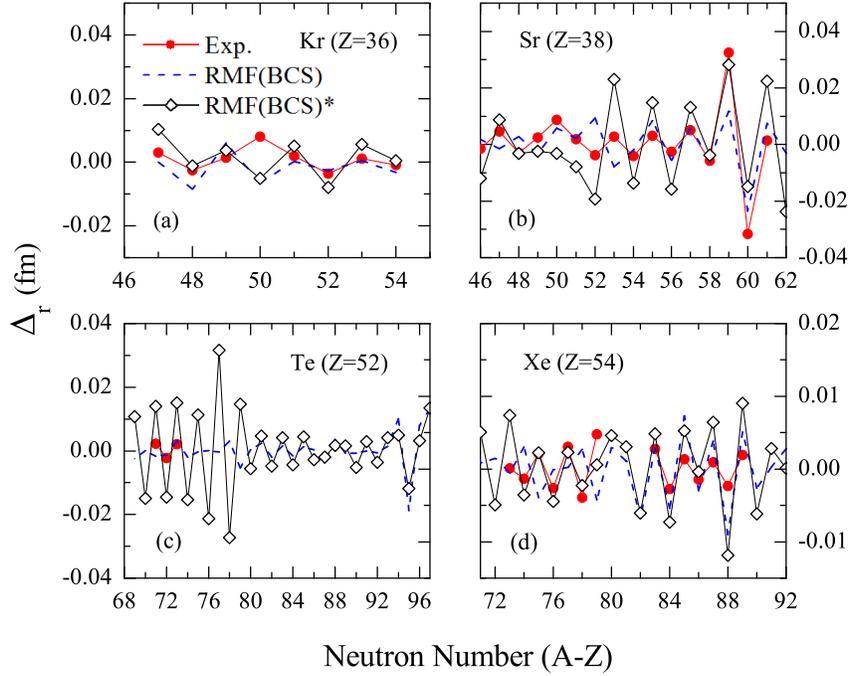}
    \caption{Odd-even staggering effects of charge radii along krypton (a), strontium (b), tellurium (c) and xenon (d) isotopic chains are obtained by the RMF(BCS) (dashed line) and the modified RMF(BCS)* (open diamond) approaches, respectively. The experimental data are taken from Ref.~\cite{ANGELI201369} (filled circle). } \label{figoe1}
\end{figure}
As shown above, the calculated results clearly show that RMF(BCS) and RMF(BCS)* approaches give almost similar trends of nuclear size along isotopic chains with $Z=36$, $38$, $52$, $54$, $56$, $58$, $60$ and $62$. Although the modified term is introduced in RMF(BCS)* approach, the quantities measuring the Cooper pair components are close for neutron and proton due to the strong coupling between different levels around Fermi surface. In order to further reflect the differences among these two methods, it is essential to investigate the odd-even variations of nuclear charge radii.
As mentioned above, the OES effects of nuclear charge radii are generally observed throughout the nuclear chart~\cite{ANGELI201369}. The possible mechanisms had been proposed, such as blocking of ground state quadrupole vibrations by the odd neutron~\cite{REEHAL1971385} and core polarization by valence neutrons~\cite{TALMI1984189,CAURIER198015}. Meanwhile, phenomenological four-particle correlations or $\alpha$-particle clustering were also supposed to produce the OES of nuclear charge radii~\cite{Zawischa:1985qds}. Another theoretical approach which includes three- or four-body part in an effective residual interaction was also introduced to reproduce the normal OES of nuclear charge radius well~\cite{ZAWISCHA1987299,PhysRevLett.61.149}. In addition, the special deformation effects also lead to the large staggering, especially in very neutron-deficient mercury and gold isotopes~\cite{GIROD19821,Ulm:1986wd}. In Ref.~\cite{weber2005effects}, it was pointed out that the size of the neutron pairing energy had an influence on the large OES of charge radii of mercury isotopes near the $N = 104$ midshell region, and the shape coexistence was also observed.

In order to emphasize these phenomena, the three-point formula has been employed to extract the local variations of charge radii along isotopic chains~\cite{Reinhard2017}. It is written in the form:
\begin{eqnarray}
\Delta_{r}(N,Z)=\frac{1}{2}[R(N-1,Z)-2R(N,Z)+R(N+1,Z)],
\end{eqnarray}
where $R(N,Z)$ is the rms charge radius of a nucleus with neutron number $N$ and proton number $Z$. In Fig.~\ref{figoe1}, odd-even staggering effects of charge radii along krypton (a), strontium (b), tellurium (c) and xenon (d) isotopic chains are shown with and without the modified term. As shown in this figure, we can find that both of these two methods can reproduce the odd-even oscillation effect. But for $^{85,86,90,91}$Sr and $^{128,129,132,133}$Xe, the OES behaviors with the RMF(BCS) approach cannot follow the trend of experimental data with respect to the RMF(BCS)* method. Meanwhile, the modified expression slightly overestimates the OES of nuclear charge radii, especially for tellurium isotopes. The same scenario can also be found along $^{90-95}$Sr isotopes. The reason is that ${D}_{\mathrm{n}}$ is larger than ${D}_{\mathrm{p}}$. For example, ${D}_{\mathrm{n}}=1.9021$ while ${D}_{\mathrm{p}}=0$ for $^{90}$Sr. As demonstrated in Ref.~\cite{An:2020qgp}, one can properly describe the OES of charge radii by adjusting the parameter $a_{0}$ in Eq.~(\ref{coop2}). In order to keep the global description, we will not further perform a fine tuning in this work.

\begin{figure}[htbp]
\includegraphics[scale=0.6]{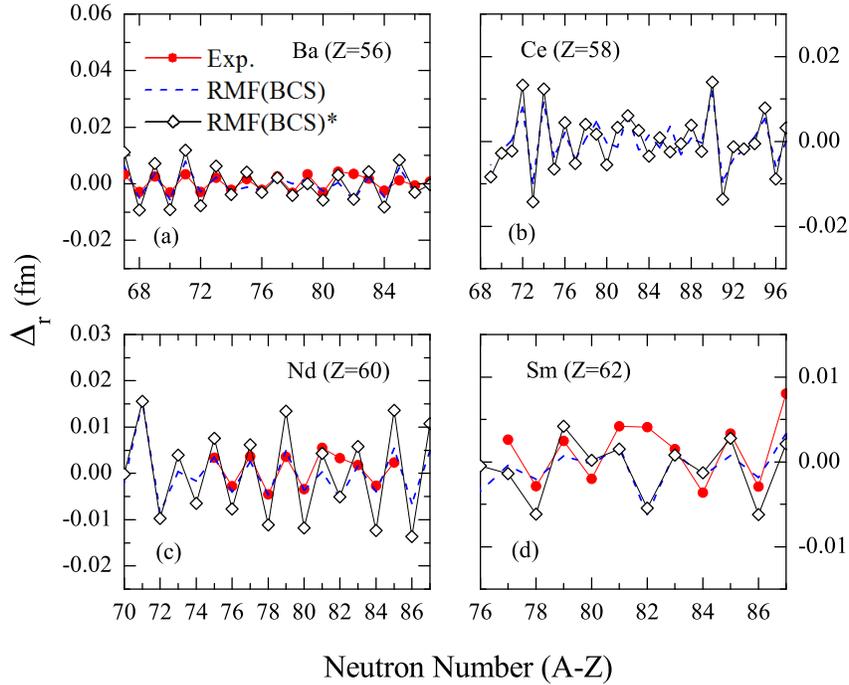}
    \caption{Same as Fig.~\ref{figoe1}, but for barium (a), cerium (b), neodymium (c) and samarium (d) isotopic chains.} \label{figoe2}
\end{figure}
In Fig.~\ref{figoe2}, the OES effects of charge radii along barium (a), cerium (b), neodymium (c) and samarium (d) isotopic chains are also shown. As encountered in tellurium isotopes, the OES behaviors are also slightly overestimated in barium and neodymium isotopic chains. In the cerium isotopic chain, the experimental results are not shown due to the absence of odd-mass nuclei data~\cite{ANGELI201369}. However, both of these two approaches give almost similar trends with the increase of neutron number except $^{142-146}$Ce isotopes. For samarium isotopes, the OES amplitudes of nuclear charge radii are also slightly overestimated by the RMF(BCS)* method. As discussed above, this is due to the overestimated difference between ${D}_{\mathrm{n}}$ and ${D}_{\mathrm{p}}$ in the modified term. As mentioned in Ref.~\cite{An:2020qgp}, this may be improved by restoring the particle number conservation in tackling pairing correlations.

The values obtained by the three-point OES formula emphasize the flattening of the isotopic dependence of the charge radii along isotopic chains~\cite{Reinhard2017}. In Fig.~\ref{figoe1} (a) and~\ref{figoe1}(b), the OES of charge radii along krypton ($Z=36$) and strontium ($Z=38$) isotopic chains cannot strictly follow the general oscillation trend at $N=50$. We definitely label this weakening behavior as ``\ abnormal staggering effect". Actually, these phenomena are also observed at neutron magic number $N=28$ (Ca), $50$ (Sr, Y, Zr), $126$ (Pb), etc~\cite{ANGELI201369}. As shown in Figs.~\ref{figoe1} and~\ref{figoe2}, similar cases are encountered at $N=82$ closed shells. This may provide a signature to identify the evidences of shell closure effects along isotopic chains in the nuclear chart. As mentioned before, nuclear charge radii may be influenced by many possible mechanisms, such as pairing interaction and quadrupole deformation, etc. Therefore, more available experimental data are needed to verify these arguments.

\section{SUMMARY AND OUTLOOK}\label{fourth}
The rapid increases of nuclear charge radii are commonly observed features across $N=50$ and $82$ closed shells throughout the periodic table~\cite{ANGELI201369}. The latest studies further demonstrate these discontinuity phenomena along cadmium and tin isotopic chains~\cite{Hammen2018,Gorges2019}. In this work, the modified formula is employed to study charge radii along rich-data even-$Z$ isotopic chains, such as krypton ($Z=36$), strontium ($Z=38$), tellurium ($Z=52$), xenon ($Z=54$), barium ($Z=56$), cerium ($Z=58$), neodymium ($Z=60$) and samarium ($Z=62$) elements. Our results can reproduce the universal kink features of nuclear charge radii, namely a smooth increase towards shell closures and then an abrupt increase through $N=50$ and $82$ filled shells. The similar scenario is encountered at $N=28$ and $126$~\cite{An:2020qgp}. From calculated results, we can find both of these methods present similar results or the modified formula shows slight improvement. This means that the $np$ pairs correction has almost no influence on charge radii for these nuclei with open proton shells and this is consistent with Ref.~\cite{Miller:2018mfb}.

 Our results can reproduce the OES behaviors of charge radii, but this trend is overestimated at magic neutron closures. Based on $\Delta_{r}$'s definition, this seemingly corresponds to the inverse OES of charge radii, namely anomalous OES behavior. As shown in Ref.~\cite{An:2020qgp}, the weakening of OES behaviors of charge radii was evidently found at $N=28$ and $126$ closed shells. Actually, this debilitating tendency can be observed naturally at neutron magic numbers~\cite{ANGELI201369}. We propose this commonly observed signature as an indicator to capture the magicity properties throughout nuclear chart. We should mention that many possible mechanisms are proposed to explain the fine structure of nuclei size~\cite{CAURIER198015,TALMI1984189,Reinhard2020PRC,Bao2020PRC,Miller:2018mfb, goodacre2020laser}, especially those about the unpaired nucleons near magicity numbers.

 The atomic nucleus is formed by two different kinds of fermions (protons and neutrons) which interact mainly by the electromagnetic and strong forces. It is pointed out that new data in neutron-rich nuclei all exhibit an intriguingly simple increase in charge radii across closed shells~\cite{GarciaRuiz:2019cog}. Moreover, the electromagnetic properties of isotopes around magic numbers of protons and neutrons have been found to exhibit astonishingly simple trends. As demonstrated in Ref.~\cite{Miller:2018mfb}, the $np$-SRCs which originate from short-range neutron-proton tensor interaction will cause protons to move far away from the center of nucleus. For open shell nuclei, the quantity of Cooper pair components coming from protons and neutrons is roughly comparable due to the strong coupling between different levels around Fermi surface. That is why both of these two approaches give almost similar results. Consider the results for both open and closed proton shell isotopes in the present and our previous works, the new ansatz is a unified approach in describing the nuclear size quantitatively. However, it is still an open question to include the $np$-SRCs self-consistently in density functional theory.

\section{Acknowledgements}
This work is supported by the Reform and Development Project of Beijing Academy of Science and Technology under Grant No. 13001-2110. This work is also supported in part by the National Natural Science Foundation of China under Grants No. 12135004, No. 11635003, No. 11961141004, No. 11025524, No. 11161130520, the National Basic Research Program of China under Grant No. 2010CB832903. X. J. is grateful for the support of the National Natural Science Foundation of China under Grants No. 11705118. L.-G. C. is grateful for the support of the National Natural Science Foundation of China under Grants No. 11975096 and the Fundamental Research Funds for the Central Universities (2020NTST06).
%This work is supported by the Reform and Development Project of Beijing Academy of Science and Technology under Grant No. 13001-2110. This work is also supported in part by the National Natural Science Foundation of China under Grants No. 11705118, No. 11975096, No. 11635003, No. 11025524, No. 11161130520, the National Basic Research Program of China under Grant No. 2010CB832903.

\bibliography{refsnew}
\end{document}